\documentclass[english,amsmath,amssymb,twocolumn,aps,prl,10pt,superscriptaddress,letterpaper,nofootinbib]{revtex4-1}

\usepackage{braket}
\usepackage{times}
\usepackage{amsmath,amssymb}
\usepackage{graphicx}
\usepackage{epsfig}
\usepackage{bm}
\usepackage{wrapfig}
\usepackage{color}

\newcommand{\code}{\texttt}

\newcommand{\hide}[1]{}

\newcommand{\eq}[1]{Eq.\,(\ref{#1})}
\newcommand{\eqs}[1]{Eqs.\,(\ref{#1})}
\newcommand{\noeq}[1]{(\ref{#1})}
\newcommand{\fig}[1]{Fig.\,\ref{#1}}

\newcommand{\nofig}[1]{(\ref{#1})}

\newcommand{\comm}[2]{\left[ #1 , #2 \right]}
\newcommand{\cum}[1]{\left\langle\!\left\langle #1 \right\rangle\!\right\rangle}
\newcommand{\erw}[1]{\left\langle\ \!\! #1 \right\rangle}

\newcommand\note[1]{\textcolor{red}{#1}}

\begin{document}

\let\jnl\rm
\def\reff@jnl#1{{\jnl#1}}

\def\physrep{\reff@jnl{Phys.~Rep.}} 

\title{Collective induced superradiant lineshifts}

\author{Gray Putnam }
\affiliation{Department of Physics, Harvard University, Cambridge, MA 02138, USA}

\author{Guin-Dar Lin}
\affiliation{National Taiwan University, Taipei, Taiwan}

\author{S. F. Yelin}
\affiliation{Department of Physics, Harvard University, Cambridge, MA 02138, USA}
\affiliation{Department of Physics, University of Connecticut, Storrs, CT 06269, USA}

\date{\today}

\begin{abstract}
Superradiant decay is accompanied by  two kinds of collective lineshifts, an induced shift and the spontaneous ``collective Lamb shift.'' Both form as sum of dipole-dipole interaction-induced level shifts between atoms in the system. We have developed a procedure to obtain numerical results on this model that self-consistently incorporates the shifts. The induced shift displays large non-zero values early in the system evolution. In addition, its effect on the superradiant system is studied: there is only a very small dephasing effect on the decay rate. While the induced shift is largely absent in not-too strongly driven systems, this parameter region might provide a good experimental regime for measuring the collective Lamb shift. These results can have important consequences for highly sensitive systems, such as quantum information science or atomic clocks.
\end{abstract}

\maketitle

\section{Introduction}

\note{
}

\begin{figure}[htbp]
\begin{center}
\includegraphics[width=0.95\linewidth]{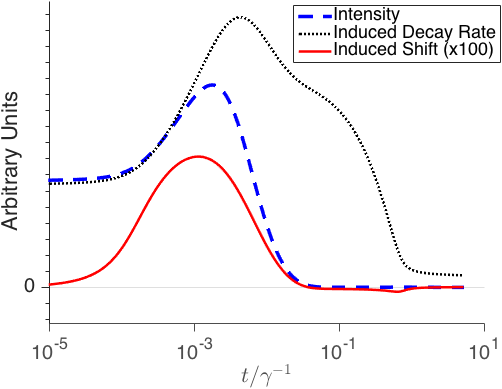}
\caption{Qualitative, logarithmic plot of superradiant evolution for a system of high optical depth. In the language of the model, the intensity is equal to the rate of change in the excited state, $-\dot{\rho}_{ee}$, the induced line shift is $\Delta$, and the decay rate is $\Gamma$. The sign convention of the induced shift is such that redshifts are positive. (The induced shift is scaled up by 100 relative to the other curves in the graph.)}
\label{fig:adot}
\end{center}
\end{figure}

Superradiance is a phenomenon where a group of particles radiate cooperatively which changes the spontaneous emission to a burst of initial, partially coherent, radiation with a long tail (see, e.g., typical example in \fig{fig:adot}). Intuitively, the particles in a superradiant system are so close together that one particle spontaneously emitting a photon is seen as phase coherent by a neighbor, who will then release a photon through stimulated emission. The coherence that is built up in this way is called "cooperative". While the ideal case of this scenario was suggested by Dicke over 60 years ago \cite{OriginalSuper} and worked on much subsequently (see, e.g., the important review \cite{haroche}), there has been a recent renewed burst of interest in the topic. This has three main reasons: First, experimental techniques have improved enough to access the more interesting parameter regimes, such as very high optical depth \cite{grimes16}, well resolved measurements in moderate density samples \cite{bromley16}, and low-excitation bad cavity regime \cite{Thompson_Nature}, among others.  Second, these cooperative effects, mostly based on dipole-dipole interactions, are strongly enhanced for long wavelengths. Thus, effects where such energy regimes could play a role, which include Rydberg transitions, vibrational and rotational molecular transitions, and even spin-flip transitions, will be strongly affected by, e.g., superradiant broadening. Third, even for optical and near-infrared transitions, let alone hyperfine transitions, such as used for atomic clocks, the accuracy of ensemble-based measurements \cite{ludlow08} has now reached a level where such collective effects will play a role. 

One aspect of superradiant decay is its associated line shift. These collective shifts have been calculated by classical means \cite{Lehmberg_Paper1,Lehmberg_Paper2}, quantum mechanically for very low \cite{Svidzinsky_Lamb,Svidzinsky_Cooperative} or first-order correlated \cite{javanainen14} or very small or thin systems \cite{adams12} and measured for moderately dense \cite{bromley16} or very lowly excited systems \cite{Thompson_Nature}. But to date, their dynamics have not been very well understood beyond the classical and toy model examples, best described in Ref.~\cite{haroche}. We will introduce in this article an analytical and numerical description of this shift, following earlier theoretical treatment of superradiant decay in Ref.~\cite{lin12}. In fact, we will show that there are two qualitatively different parts to the shift, one that is induced and thus visible only for those systems and times where strong decay is present and one that is spontaneous, the so-called ``collective Lamb'' shift whose vacuum limit is the traditional Lamb shift. This collective Lamb shift can be measured best in situations where strong superradiant decay is not present. It is, in particular, interesting to derive the parametric dependence of this shift: as opposed to the strength of radiative decay which depends on the optical depth of the sample, the shift depends on the density per cubic wavelength. 

When introducing the analytical form of collective line shifts we will present a brief qualitative review of our earlier analytical derivations and then discuss the explicit form and numerical result of the most generic appearance of collective induced shifts in particular. 

\section{Model}
\label{model}

We start with the master equation derived in Ref.~\cite{lin12}:
\begin{eqnarray}
\label{eq:master}
\lefteqn{ 
\dot\rho^{(1,2)}(t) \;=\; }\\
&& i \! \sum_{i=1,2} \left[ \sigma_{i}\Omega^\dagger_{Li} + \sigma^+_{i} \Omega_{Li}\,,\, \rho^{(1,2)} \right]  \nonumber \\
&& - i \! \sum_{i=1,2} \Delta_{ii} \comm{\rho^{(1,2)}}{\comm{\sigma_i^+}{\sigma_i}}
-i \! \!\! \sum_{i,j=1,2}\delta_{ij}\comm{\rho^{(1,2)}}{\sigma_j^+\sigma_i} \nonumber\\
&& -  \sum_{i,j=1,2} \bigg( \frac{\Gamma_{ij}}{2} 
\left( \left[\rho^{(1,2)}\sigma_{i},\sigma^+_{j}\right] + \left[ \sigma_{i},\sigma^+_{j}\rho^{(1,2)}\right] \right) \nonumber\\
&& \qquad+ \frac{\Gamma_{ij}+\gamma_{ij}}{2} 
\left( \left[\rho^{(1,2)}\sigma^+_{j},\sigma_{i}\right] + \left[ \sigma^+_{j},\sigma_{i}\rho^{(1,2)}\right] \right) \bigg). \nonumber
\end{eqnarray}
This describes the dynamics of a two-atom two-level density matrix of some (randomly picked) two atoms labeled 1 and 2 with excited states $\ket{e_i}$ and ground states $\ket{g_i}$. Here, $\Omega_{Li}$ is the driving field including the Lorentz-Lorenz local field correction at the location of atom $i$, and the atom $i$'s lowering operator is  $\sigma_i = \ket{g_i}\bra{e_i}$. Most importantly, $\gamma_{ij}$/$\Gamma_{ij}$ and $\delta_{ij}$/$\Delta_{ij}$ denote the collective spontaneous/induced decay and frequency shift parameters (i.e., $\delta_{ij}$ would be the ``collective Lamb shift''). For a single atom in vacuum, the spontaneous decay tends toward the vacuum decay rate $\gamma_0$ for the $e\rightarrow g$ transition of the atoms and the spontaneous shift toward the usual Lamb shift. All of those four parameters result from second-order correlation of the dense atomic sample: 
\begin{eqnarray}
\label{eq:gammadelta}
\frac{\Gamma_{ij}
}{2}-i\Delta_{ij}
&=& \frac{\wp^2}{\hbar^2} \int\limits_0^\infty \! d\tau  \cum{E_i^-(t)\,E_j^+(t-\tau)} e^{-i\omega\tau} ,\\
\frac{\gamma_{ij}}{2} -  i\,\delta_{ij} &=&
\frac{\wp^2}{\hbar^2} \int\limits_0^\infty\!  d\tau \erw{ \left[ E_i^+(t-\tau),E_j^-(t)\right]} e^{-i\omega\tau}. \nonumber
\end{eqnarray}
Here, $\cum{\ldots}$ denotes the cumulant, $\wp$ the transition dipole matrix element, and $E^\pm_i$ the positive/negative frequency part of the electric field operator at the location of atom $i$. The form of these parameters was found using a Schwinger-Keldysh formalism up to second order in the field-field correlations\footnote{Note that this second-order approximation is very good for typical superradiant systems, but would break down for strongly entangled systems.}. The assumptions leading to this form of the equations are a Markovian radiation reservoir (i.e., free space) and a three-dimensional homogeneous atomic gas. Thus, this is a very good description for macroscopic atomic gases, but not applicable for lower dimensions or ordered systems. It should be noted that the projection of \eqs{eq:gammadelta} into coordinate space would result in the usual full complex form of the dipole-dipole interactions. 

Why are the decay and shift quantities called ``induced'' vs.\ ``spontaneous''? While for the full calculation the quantum field has to be treated in a multi-mode fashion, a good intuitive understanding can be gained from the following (ignoring time dependence):  $E^- \propto \hat a$, $E^+\propto\hat a^\dagger$, thus $\cum{E_i^-(t)\,E_j^+(t-\tau)} \propto \erw{\hat a \hat a^\dagger} \approx n$ for photon number $n$, thus, this describes a light-induced quantity, while $\erw{ \left[ E_i^+(t-\tau),E_j^-(t)\right]} \propto \erw{\comm{\hat a}{\hat a^\dagger}} = {\cal O}(1)$, i.e., this describes a spontaneous (i.e., photon-number independent) quantity.

The exact form of the $\Gamma$/$\Delta$ and $\gamma$/$\delta$ pairs can now be calculated, for example, using a self-consistent Hartree approximation to be solved using a Dyson equation. This has been done in Ref.~\cite{lin12} for $\Gamma$ where the other parameters were approximated to their vacuum values. Here, we will also neglect the collective modification of the two spontaneous parameters, $\gamma$ and $\delta$. For the case of highly excited systems, such as the traditional superradiant case, they are of the order or smaller than the vacuum values. Their significance in, e.g., driven systems will be discussed elsewhere. 

Here is a brief outline of how to determine quantitative values for $\Gamma, \Delta$. Equation \noeq{eq:gammadelta} can be regarded as an  equation in variables $\rho^{(1,2)}_{\alpha\beta}$, $\Gamma_{ij}$, and $\Delta_{ij}$.  The calculation then consists of the following steps: (i) Assume that we can neglect retardation, i.e., that the system is small enough that the time for light to travel through the system is small compared to system evolution times.  Thus, $\rho_{\alpha\beta}$ is independent of the choice of the two probe atoms, and $\Gamma$ and $\Delta$ are only dependent on whether $i=j$ (in this case: $\Gamma_{ii}\equiv \Gamma$, $\Delta_{ii}\equiv \Delta$) or $i\ne j$ (in this case $\Gamma_{ij}\equiv \bar\Gamma$ and $\Delta_{ij}\equiv\bar\Delta$). This leaves, explicitly, the following parts of the density operator: average excited state population $a\equiv \overline{{\rm Tr}(\sigma_i^+\sigma_i\rho)}$, average effective two-atom inversion $n \equiv\overline{{\rm Tr}(\sigma^z_1\sigma^z_2\rho)}$, average two-atom cross correlation $\rho_{eg,ge}\equiv\overline{{\rm Tr}(\sigma_1\sigma_2^+\rho)}$, and average single-atom coherence $\rho_{eg}\equiv\overline{{\rm Tr}(\sigma_i\rho)}$. (ii) Use a self-consistent Dyson equation to determine $\Gamma,\bar\Gamma, \Delta, \bar\Delta$ as functions of the density matrix and each other. This can now been solved explicitly numerically. (For exact formulas, see SOM.) (iii) While $\Gamma$ and $\bar\Gamma$ can be approximately solved analytically (see SOM), $\Delta$ can only be solved numerically by explicitly calculating the Kramers-Kronig form
\begin{equation}
\label{eq:defdelta}
\Delta \;=\; \frac{1}{\pi} {\cal P}\int\limits_{-\infty}^\infty\!d\Delta' \frac{\Gamma(\Delta')}{\Delta-\Delta'},
\end{equation}
where ${\cal P}$ denotes the principal part of the integral. 

While steps (i) and (ii) have been done explicitly in Ref.~\cite{lin12}, we show the results of part (iii) here. In particular, we are going to discuss the following questions (and answers) regarding the outcome of those calculations: 
\begin{itemize}
\item Can one get $\Delta\ne 0$ for an (approximately) spherical volume if atoms are not a priori polarized? $\longrightarrow$ Yes.
\item If the value of $\Delta$ is finite, is it big enough to be measurable? $\longrightarrow$ Yes, in principle.
\item What is the time dependence during superradiant emission? $\longrightarrow$ Chirp from red to blue.
\item Which macroscopic parameters does $\Delta$ depend on? In particular, while some references give the macroscopic observables as depending on the relative density ${\cal N}\lambda^3$, we \cite{lin12} (along with other authors e.g. in Ref.~\cite{akkermans08}) find that $\Gamma$ depends nearly exclusively on the optical depth ${\cal N} \lambda^2\ell$.  $\longrightarrow$ $\Delta$ (as an intensive variable) depends on the relative density.
\end{itemize}

\section{Results for collective induced lineshifts in a superradiant system}

In this section, we describe the results of the numerical calculation of the induced shift $\Delta$ and related quantities. As the simplest generic example, all the presented results assume an initially completely inverted two-level sample without driving field. 

For all results, the system evolution time is defined in terms of  the inverse vacuum decay rate $\gamma^{-1}$, the relative density is given in terms of the cooperativity parameter ${\cal C}={\cal N}\lambda^3/4\pi^2$, the relative system size $\ell$ is proportional to the propagation length in the medium $l$ compared to the wavelength, $\ell=\pi l/\lambda$. Thus, the optical depth $\eta={\cal N}\lambda^2 l \propto {\cal C}\ell$. These quantities are used as parameters for the following graphs.

\subsection{Numerical Methods}
\label{sec:numerics}

We used an adaptive 4th-5th order Runge Kutta algorithm \cite{runge_kutta_history} to integrate the master equation \noeq{eq:master}.   $\Gamma$ and $\overline{\Gamma}$ were defined using \eq{eq:gammadelta} and $\Delta$ was found from \eq{eq:defdelta}. The principal value integration of the Kramers-Kronig relation was approximated using the results of \cite{nyiri2000numerical}. At this point in the calculation, the parameters $\Gamma , \Delta$ were still defined in terms of each other, so a 2-D least squares method (implemented by the Matlab function \code{lsqnonlin}) was used to extricate the values of $\Gamma$ and $\Delta$ at each time step. These two values could then be plugged in to calculate $\bar{\Gamma}$ and in turn the master equation. This procedure allowed us to obtain results on the evolution of a superradiant system for given parameter values.

\subsection{Results for $\Gamma$ and $\Delta$}

\begin{figure}[htbp]
\begin{center}
\includegraphics[scale=.43]{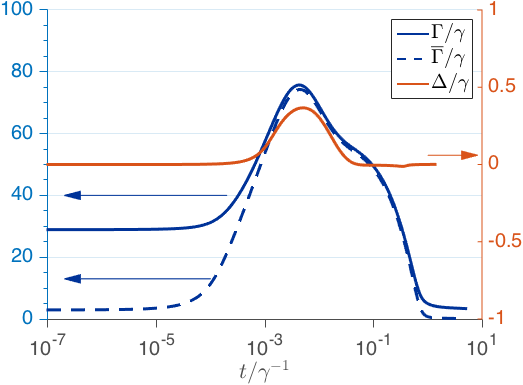}\\
\includegraphics[scale=.43]{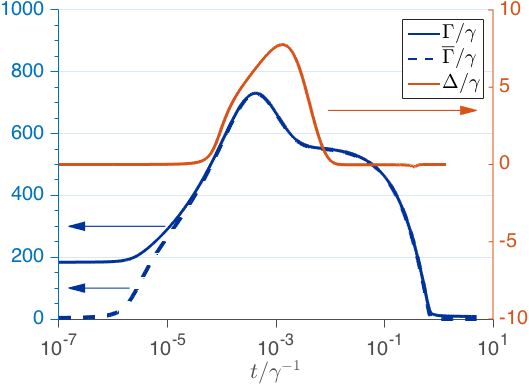}
\caption{(upper) Plot of $\Gamma$ and $\bar{\Gamma}$ (left axis) and $\Delta$ (right axis) for logarithmic time in terms of the vacuum decay rate $\gamma$ for  $\mathcal{C} = 10$ and $\ell = 10$. The sign convention of $\Delta$ is such that a redshift is positive. 
(lower) Same plot, but for $\mathcal{C} = 31$,  $\ell = 31$. The system first undergoes a chirp towards the red for the ``superradiant" evolution (as shown by the large and short-lived peak in $\Delta$) then goes to blue (as shown by the small and long-lived dip in $\Delta$).}
\label{fig:pic2}
\end{center}
\end{figure}

\begin{figure}[htbp]
\begin{center}
\includegraphics[scale=.43]{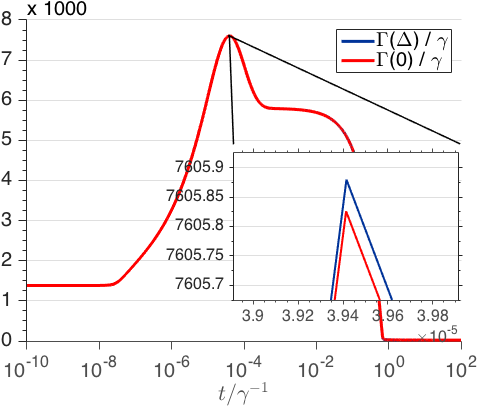}
\caption{Plot of $\Gamma$ for $\Delta$ set to 0 for the entire evolution (red) and $\Delta$ found self-consistently (blue) for $\mathcal{C} = 100$ and $\varrho = 100$. This plot proves that there is nearly no difference for typical parameter choices.}
\label{fig:GammaEffect}
\end{center}
\end{figure}

The numerical results allow us to obtain understanding of how the line shift of a superradiant system changes with time. Figure \nofig{fig:pic2} shows the typical time dependence for two different optical depths. The most important features are: (i) The system undergoes a chirp, first towards the red for the main ``superradiant'' evolution, then to the blue. The redshift is strong and very short, the blue shift long and very weak. The qualitative red-to-blue shift has been predicted in a very simple model in Ref.~\cite{haroche}. For an (approximately) spherical volume one could naively argue that no shift could be seen, since the dipole-dipole shifts would be expected to cancel each other. Obviously, this has to be true on a time average - but the Kramers-Kronig relation between $\Gamma$ and $\Delta$ would make $\Delta\equiv 0$ unphysical. (ii) While for typical systems the shift can be of the order or larger than the natural line width of the transition, the superradiant decay rate (and thus, the minimum possible line broadening) are orders of magnitude larger. (iii) In \fig{fig:GammaEffect}, the broadening/decay for an assumed zero shift is compared to the realistic non-zero shift. It can be clearly seen that for typical situations, replacing $\Gamma(\Delta=0)$ for $\Gamma(\Delta)$ can be expected to be a very good approximation. 



\subsection{Density dependence of $\Delta$}

\begin{figure}[h]
\begin{center}
\includegraphics[scale=.43]{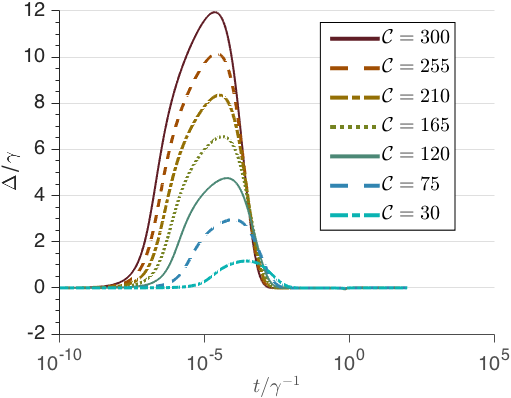}\\
%
\includegraphics[scale=.43]{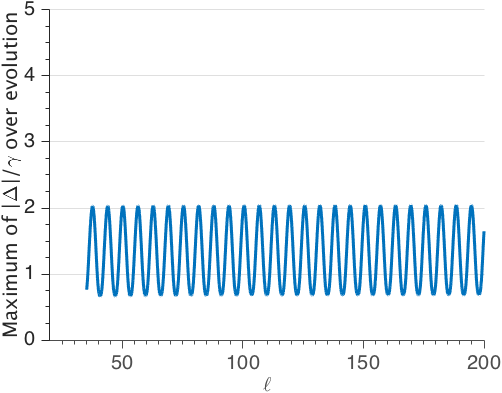}
\caption{(upper) Plots of $\Delta$ for logarithmic time in terms of $\gamma$ with $\varrho = 10$,  and $\mathcal{C}$ varied. The sign convention of $\Delta$ is such that a redshift is positive. (lower) The maximum redshift value of $\Delta$ over the evolution for different values of $\ell$. ${\cal C} = 20$ for all points. The plot has the expected periodicity of $\ell \approx\pi$.}
\label{fig:varyCr}
\end{center}
\end{figure}

One of the persistent questions of this field is whether superradiance depends on the optical depth or the relative density. As was found earlier, the magnitude of the superradiant decay (and thus the resulting radiated intensity) depends dominantly on the optical depth. The question is thus whether this is also the case for the induced shift. Since there are arguments for both the optical depth (everything else depends on the optical depth only) as well as for the relative density (the optical depth is an extensive variable, but the shift can be expected to be intensive, as is the relative density), the best test is numerical. The result is given in \fig{fig:varyCr}: The dependence of the system length $\ell$ only is trivially oscillatory around a constant value (not surprising because of the sinusoidal terms in the dipole-dipole interaction), the maximum, and thus the strength, of $\Delta$ does depend strongly on the relative density. 

\section{Driven system}
Another important cooperative regime is a driven, stationary system. This system potentially displays a large array of interesting effects. It is also a very good regime to study the collective Lamb shift. One would expect that the main contribution to the induced shift in this case would come from the second order in the driving field (this is typically the main contribution to upper-state population). It turns out, however that to this order $\Delta$ is actually identically equal to zero. Without going into the mathematical detail, here is a short outline of why this is the case: The induced parameters, $\Gamma, \bar\Gamma$, and $\Delta$ are all, via the Dyson equation, proportional to the atom source functions, 
\begin{eqnarray*}
\lefteqn{P^{s}_{ij}(t) \;=}\\
&=& \frac{\wp^2}{\hbar^2} {\cal N}^2 \bigg( \cum{\tilde\sigma_i^+(t),\sigma_j(t)} + \cum{\sigma_i^+(t), \tilde\sigma_j(t)} \bigg),
\end{eqnarray*}
where we find, up to the second order in the driving field, $P^{(s)}_{ii} \propto a^{(2)}-\left|\rho_{eg}\right|^2$ and $P^{(s)}_{i\ne j} \propto x^{(2)}-\left|\rho_{eg}\right|^2$, where the superscript gives the order in the driving field. Both of these terms are identically equal to zero, and thus, so are all the induced parameters. This is in excellent agreement with measurements, such as \cite{bromley16}. 

While these terms are expected to give contributions for higher orders (i.e., stronger driving), we will not discuss this case here. What is important to note, however, that this regime would be the ideal regime to measure the collective spontaneous modification of the decay and the collective Lamb shift, since they are not overshadowed by the induced quantities in this case.


\section{Conclusion}


We have summarized, in this article, analytical evidence and numerical results for the elusive ``collective'' (or ``cooperative'') shift. Moreover, we have identified parameter regimes where the induced shift can be measured and where the Lamb shift dominates -- and experimental evidence, as far as it exists to date \cite{grimes16,bromley16} backs up our result. It is important to understand the difference between the two: while, intuitively, the induced shift can be, for example, understood as a Stark-type shift that results from the presence of cooperative radiation in the system, the collective Lamb shift is modified from its vacuum counterpart due to the different distribution of vacuum modes that exist in such a sample. Accordingly, time and excitation-level dependence of the two shifts are very different from each other. 

While there is considerable fundamental interest in the cooperative shifts, understanding them is of particular practical value for ensemble-based metrology. Because clocks, in particular atomic clocks based on ensembles, depend on having very narrow and very accurate lines, even the small cooperative shifts that result in moderately dense media can, by now, become in principle a dominant contribution to measurement uncertainties. This is an important question also for high-coherence applications such as quantum information science, especially for long-wavelength transitions such as Rydberg transitions, molecular vibrational or rotational transitions, or even spin-flip transitions, since the main parameters, the relative density grows with the wavelength to the third order. 

\bibliographystyle{apsrev4-1}
\bibliography{gray_draft}

\end{document}